\documentclass[12pt]{article}
\usepackage{graphicx}

\newcommand{\bc}{bchain}

\begin{document}

\newenvironment{describe}{\begin{list}{}{\setlength\leftmargin{80pt}}\setlength\labelsep{10pt}\setlength\labelwidth{70pt}}{\end{list}}

\newenvironment{flag}{\begin{list}{\makebox[20pt]{\hss$\circ$\enspace}}
                                  {\labelwidth 20pt}}{\end{list}}



\newenvironment{numberedlist}
{\begin{list}{\makebox[20pt]{\hss(\arabic{itemno})\enspace}}
             {\usecounter{itemno}\labelwidth 20pt}}{\end{list}}

\newenvironment{alphabetlist}
{\begin{list}{\makebox[20pt]{\hss(\alph{itemno1})\enspace}}
             {\usecounter{itemno1}\labelwidth 20pt}}{\end{list}}

\newenvironment{romanlist}
{\begin{list}{\makebox[20pt]{\hss(\roman{itemno2})\enspace}}
             {\usecounter{itemno2}\labelwidth 20pt}}{\end{list}}

\newcounter{itemno}

\newcounter{itemno1}

\newcounter{itemno2}
\newcounter{lemma}
\newcounter{exno}

\newcounter{defno}







\newenvironment{defn}{\refstepcounter{defno}\medskip \noindent {\bf
Definition \thedefno.\ }}{\medskip}

\newenvironment{ex}{\refstepcounter{exno}\medskip \noindent {\bf
Example \theexno.\ }}{\medskip}

\newenvironment{millerexample}{
 \begingroup \begin{tabbing} \hspace{2em}\= \hspace{5em}\= \hspace{5em}\=
\hspace{5em}\= \kill}{
 \end{tabbing}\endgroup}

\newenvironment{wideexample}{
 \begingroup \begin{tabbing} \hspace{2em}\= \hspace{10em}\= \hspace{10em}\=
\hspace{10em}\= \kill}{
 \end{tabbing}\endgroup}

\newcommand{\sep}{\;\vert\;}

\newcommand{\ra}{\rightarrow}
\newcommand{\app}{\ }
\newcommand{\appt}{\ }
\newcommand{\tup}[1]{\langle\nobreak#1\nobreak\rangle}

\newcommand{\hu}{{\cal H}^+}
\newcommand{\Free}{{\cal F}}
\newcommand{\oprove}{\vdash\kern-.6em\lower.7ex\hbox{$\scriptstyle O$}\,}
\newcommand{\true}{\top}

\newcommand{\Dscr}{{\cal D}}
\newcommand{\Pscr}{{\cal P}}
\newcommand{\Gscr}{{\cal G}}
\newcommand{\Fscr}{{\cal F}}
\newcommand{\Vscr}{{\cal V}}
\newcommand{\Uscr}{{\cal U}}
\newcommand{\pderivation}{{\cal P}\kern -.1em\hbox{\rm -derivation}}
\newcommand{\pderivationl}{{\cal P}\kern -.1em\hbox{\em -derivation}}
\newcommand{\pderivable}{{\cal P}\kern -.1em\hbox{\rm -derivable}}
\newcommand{\pderivablel}{{\cal P}\kern -.1em\hbox{\em -derivable}}
\newcommand{\pderivations}{{\cal P}\kern -.1em\hbox{\rm -derivations}}
\newcommand{\pderivability}{{\cal P}\kern -.1em\hbox{\rm -derivability}}
\newcommand{\eqm}[1]{=_{\scriptscriptstyle #1}}
\newcommand\subsl{\preceq}
\newcommand{\fnrestr}{\uparrow}

\newcommand{\match}{{\rm MATCH}}
\newcommand{\triv}{{\rm TRIV}}
\newcommand{\imit}{{\rm IMIT}}
\newcommand{\proj}{{\rm PROJ}}
\newcommand{\simpl}{{\rm SIMPL}}
\newcommand{\failed}{{\bf F}}

\newcommand{\Dsiginst}[1]{{[#1]_\Sigma}}
\newcommand{\Psiginst}[1]{{[#1]_\Sigma}}
\newcommand{\lnorm}{{\lambda}norm}
\newcommand{\seq}[2]{#1 \supset #2}
\newcommand{\dseq}[2]{#1_1,\ldots,#1_{#2}}

\newcommand{\all}{\forall}
\newcommand{\some}{\exists}
\newcommand{\lambdax}[1]{\lambda #1\,}
\newcommand{\somex}[1]{\some#1\,}
\newcommand\allx[1]{\all#1\,}

\newcommand{\subs}[3]{[#1/#2]#3}
\newcommand{\rep}[3]{S^{#2}_{#1}{#3}}
\newcommand{\ie}{{\em i.e.}}
\newcommand{\eg}{{\em e.g.}}

\newcommand{\lbotr}{$\bot$-R}
\newcommand{\ldbotr}{\bot\mbox{\rm -R}}
\newcommand{\landl}{$\land$-L}
\newcommand{\ldandl}{\land\mbox{\rm -L}}
\newcommand{\landr}{$\land$-R}
\newcommand{\ldandr}{\land\mbox{\rm -R}}
\newcommand{\lorl}{$\lor$-L}
\newcommand{\ldorl}{\lor\mbox{\rm -L}}
\newcommand{\lorr}{$\lor$-R}
\newcommand{\ldorr}{\lor\mbox{\rm -R}}
\newcommand{\limpl}{$\supset$-L}
\newcommand{\ldimpl}{\supset\mbox{\rm -L}}
\newcommand{\limpr}{$\supset$-R}
\newcommand{\ldimpr}{\supset\mbox{\rm -R}}
\newcommand{\lnegl}{$\neg$-L}
\newcommand{\ldnegl}{\neg\mbox{\rm -L}}
\newcommand{\ldnegr}{\neg\mbox{\rm -R}}
\newcommand{\lalll}{$\forall$-L}
\newcommand{\ldalll}{\forall\mbox{\rm -L}}
\newcommand{\lallr}{$\forall$-R}
\newcommand{\ldallr}{\forall\mbox{\rm -R}}
\newcommand{\lsomel}{$\exists$-L}
\newcommand{\ldsomel}{\exists\mbox{\rm -L}}
\newcommand{\lsomer}{$\exists$-R}
\newcommand{\ldsomer}{\exists\mbox{\rm -R}}
\newcommand{\ldlamlr}{\lambda}
\newcommand{\sequent}[2]{\hbox{{$#1\ \longrightarrow\ #2$}}}
\newcommand{\prog}[2]{\hbox{{$#1\ \supset\ #2$}}}
\newcommand{\run}{\Gamma}

\newcommand{\Ibf}{{\bf I}}
\newcommand{\Cbf}{{\bf C}} 
\newcommand{\Cbfpr}{{\bf C'}}

\newcommand{\cprove}{\vdash_C}
\newcommand{\iprove}{\vdash_I}

\newsavebox{\lpartfig}
\newsavebox{\rpartfig}


\newenvironment{exmple}{
 \begingroup \begin{tabbing} \hspace{2em}\= \hspace{3em}\= \hspace{3em}\=
\hspace{3em}\= \hspace{3em}\= \hspace{3em}\= \kill}{
 \end{tabbing}\endgroup}
\newenvironment{example2}{
 \begingroup \begin{tabbing} \hspace{8em}\= \hspace{2em}\= \hspace{2em}\=
\hspace{10em}\= \hspace{2em}\= \hspace{2em}\= \hspace{2em}\= \kill}{
 \end{tabbing}\endgroup}

\newenvironment{example}{
\begingroup  \begin{tabbing} \hspace{2em}\= \hspace{3em}\= \hspace{3em}\=
\hspace{3em}\= \hspace{3em}\= \hspace{3em}\= \hspace{3em}\= \hspace{3em}\= 
\hspace{3em}\= \hspace{3em}\= \hspace{3em}\= \hspace{3em}\= \kill}{
 \end{tabbing} \endgroup }

\newcommand{\sand}{sand} 
\newcommand{\pand}{pand} 
\newcommand{\cor}{cor} 

\newcommand{\lb}{\langle}
\newcommand{\rb}{\rangle}
\newcommand{\pr}{prov}
\newcommand{\prG}{intp}
\newcommand{\prSG}{intp_E}
\newcommand{\intp}{intp_o}
\newcommand{\prove}{exec} 
\newcommand{\np}{invalid} 
\newcommand{\Ra}{\supset}  
\newcommand{\add}{\oplus} 
\newcommand{\adc}{\&} 
\newcommand{\Cscr}{{\cal C}}
\newcommand{\seqweb}{SProlog}
\newcommand{\sprog}{{SProlog}}

\newtheorem{theorem}[lemma]{Theorem}

\newtheorem{proposition}[lemma]{Proposition}

\newtheorem{corollary}[lemma]{Corollary}
\newenvironment{proof}
     {\begin{trivlist}\item[]{\it Proof. }}%
     {\\* \hspace*{\fill} \end{trivlist}}

\newcommand{\seqand}{\prec}
\newcommand{\seqor}{\cup}
\newcommand{\seqandq}[2]{\prec_{#1}^{#2}}
\newcommand{\parandq}[2]{\land_{#1}^{#2}}
\newcommand{\exq}[2]{\exists_{#1}^{#2}}
\newcommand{\ext}{intp_G}

\title{\bf  Mutually Exclusive Modules  in  Logic Programming }
\author{
 Keehang Kwon\\
\sl \small Faculty of Computer Eng., DongA  University\\
\small khkwon@dau.ac.kr
}
\maketitle



\newcommand{\prov}{ex}

\noindent {\bf Abstract}: 
Logic programming has traditionally lacked devices for expressing 
 mutually exclusive modules. We address this limitation 
  by adopting 
   choice-conjunctive modules of the form 
 $D_0 \adc D_1$  where  $D_0, D_1$ are a conjunction of Horn clauses and 
 $\adc$ is a linear logic connective.   Solving a  goal $G$ using  $D_0 \adc D_1$ -- $\prov(D_0 \adc D_1,G)$ -- has the 
 following operational semantics: $choose$ a successful one between $\prov(D_0,G)$ and
 $\prov(D_1,G)$. In other words, if $D_0$ is chosen in the course of solving $G$, then $D_1$ will be
discarded and vice versa.  Hence, the class of  choice-conjunctive  modules can capture
the notion  of  mutually exclusive modules.

{\bf keywords:} mutual exclusion, cut, linear logic, choice-conjunction.

\maketitle

\newcommand{\muprolog}{LProlog}


\section{Introduction}\label{sec:intro}

Modern logic programming languages support a notion of modules, \ie,  a conjunction of clauses as a unit.
Despite their attractiveness, logic programming has 
traditionally lacked elegant devices for structuring mutually exclusion at the module level.
 Lacking such devices, structuring mutually exclusive modules in logic programming has  been impossible.

\newcommand{\hweb}{MutexWeb}

This paper proposes a logical, high-level solution to this problem. 
To be specific, we propose
\hweb, an extension to LogicWeb with a novel feature called choice-conjunctive modules.
 This logic extends  modules  by the choice construct 
 of the form $D_0 \adc D_1$ where $D_0, D_1$ are modules and $\adc$ is a choice-conjunctive connective of
linear logic.
Inspired by  \cite{Jap03}, this has the following intended semantics: $choose$ a successful one between $D_0$ and $D_1$ in the course of 
solving a goal. Of course, the unchosen module will be discarded.
This expression thus supports the idea of mutual exclusion. 

An illustration of this aspect is provided by the following modules $quicksort, heapsort$ which define the 
usual
$qsort, hsort$ relation:

\begin{exmple}
$mod(quicksort).$  \% quicksort \\ 
$ qsort(X,L)  :-  \ldots. $ \\
$\vdots$ \\ \\ \\
$mod(heapsort).$   \% heapsort  \\
$ hsort(X,L)  :- \ldots .$ \\
$\vdots$
\end{exmple}
\noindent 
Now we want to define a module $sort$ which contains different sorting algorithms.
This is show below:

\begin{exmple}
$mod(sort).$  \%  modue sort \\ 
$mod(quicksort)\  \adc\ mod(heapsort). $ \\
\end{exmple}
\noindent In the above, these two sorting algorithms
 are defined as  mutually exclusive. Hence, only one of these two sorting algorithms
can be used. 

The remainder of this paper is structured as follows. We describe \hweb\
 in
the next section. In Section \ref{sec:modules}, we
present some examples of \hweb.
Section~\ref{sec:conc} concludes the paper.

\section{The Language}\label{sec:logic}

The language is an extended  version of Horn clauses
 with choice-conjunctive modules and implication goals. It is described
by $G$- and $D$-formulas given by the syntax rules below:
\begin{exmple}
\>$G ::=$ \>   $A \sep  G \land  G \sep   D \Ra  G \sep \some x\ G $ \\   \\
\>$D ::=$ \>  $A   \sep  G \supset D\ \sep \all x\ D \sep D \adc D \sep  D \land D $\\
\end{exmple}
\noindent
In the rules above,   
$A$  represents an atomic formula.
A $D$-formula  is called a  
   module. 
 
In the transition system to be considered, $G$-formulas will function as 
queries and a set of $D$-formulas will constitute  a program. 

 We will  present an operational 
semantics for this language. The rules of \hweb\ are formalized by means of what it means to
execute a goal task $G$ from a program $\Pscr$.
These rules in fact depend on the top-level 
constructor in the expression,  a property known as
uniform provability\cite{Mil89jlp,MNPS91}. Below the notation $\bc(D,\Pscr,A)$ denotes
that the $D$ formula is distinguished
(marked for backchaining). Note that execution  alternates between 
two phases: the goal-reduction phase (one  without a distinguished clause)
and the backchaining phase (one with a distinguished clause).

\begin{defn}\label{def:semantics}
Let $G$ be a goal and let $\Pscr$ be a program.
Then the notion of   executing $\lb \Pscr,G\rb$ -- $\prov(\Pscr,G)$ -- 
 is defined as follows:
\begin{numberedlist}

\item  $\bc(A,\Pscr,A)$. \% This is a success.

\item    $\bc(G_1\supset D,\Pscr,A)$ if 
 $\prov(\Pscr, G_1)$ and  $\bc(D,\Pscr, A)$.

\item    $\bc(\all x D,\Pscr,A)$ if   $\bc([t/x]D,\Pscr, A)$.

\item    $\bc(D_0 \land D_1,\Pscr,A)$ if   $\bc(D_0,\Pscr, A)$.

\item    $\bc(D_0 \land D_1,\Pscr,A)$ if   $\bc(D_1,\Pscr, A)$.

\item    $\bc(D_0 \adc D_1,\Pscr,A)$ if  choose a successful disjunct between  $\bc(D_0,\Pscr, A)$ and
 $\bc(D_1,\Pscr, A)$. 

\item    $\prov(\Pscr,A)$ if   $D \in \Pscr$ and $\bc(D,\Pscr, A)$. \%  change to backchaining phase.

\item  $\prov(\Pscr,G_1 \land G_2)$  if $\prov(\Pscr,G_1)$  and
  $\prov(\Pscr,G_2)$.

\item $\prov(\Pscr,\exists x G_1)$  if $\prov(\Pscr,[t/x]G_1)$.

\item $\prov(\Pscr, D\Ra G_1)$ if $\prov(\{ D \}\cup \Pscr,G_1)$

\end{numberedlist}
\end{defn}

\noindent  
In the rule (6), the symbol $D_0 \adc D_1$  allows for the mutually exclusive execution of modules. This rule
 can be implemented as follows:
  first attempts to solve the goal using $D_0$.
 If it succeeds, then do nothing (and do not leave any choice point for $D_1$
). If it fails, then $D_1$ is attempted. 

Our execution model based on uniform proof is not complete with respect to linear logic.
However, it is complete with respect to affine logic (linear logic + weakening).
The following theorem connects our language to affine logic.
Its proof can be obtained from the fact that the cut rule is admissible in affine logic.

\begin{theorem}
 Let $\{ D_1,\ldots,D_n \}$ be a program and 
let $G$ be a goal.  Then, $\prov(\{ D_1,\ldots,D_n \},G)$ terminates with a success
 if and only if $G$ follows from
$\{ ! D_1,\ldots, ! D_n \}$ in intuitionistic affine logic. 
\end{theorem}
\noindent In the above, $! D$ represents that $D$ is  a reusable clause.

\section{\hweb}\label{sec:modules}

In our context, a web page corresponds simply to a set of $D$-formulas
 with a URL. 
The module construct $mod$ allows a URL to be associated to a set of $D$-formulas.
An example of the use of this construct is provided by the 
following ``lists'' module which contains some basic list-handling rules.

\begin{exmple}
 $mod(lists)$.\\
\%  deterministic version of the member predicate \\
$memb(X,[X|L])\ \adc $\\
$memb(X,[Y|L])$ {\rm :-}  $(neq\ X\ Y)\ \land\ memb(X,L).$\\
\%  optimized version of the append predicate \\
$      append([],L,L)\ \adc $ \\
$      append([X|L_1],L_2,[X|L_3])$ {\rm :-} $append(L_1,L_2,L_3).$  \\ 

\% the sorting of a list via two mutually exclusive sorting algorithms \\
$mod(quicksort)\ \adc\ mod(heapsort)$\\
\end{exmple}
\noindent Our language  makes it possible to use quicksort and heapsort in a mutually exclusive way.

 These  pages can be made available in specific contexts by explicitly
mentioning the module implication. For example, consider a goal 
$mod(lists) \Ra  qsort([2,60,3,5],L)$. 
Solving this goal  has the effect
 of adding the rules in $lists$
to the program before evaluating $qsort([2,60,3,5],L)$, producing the result $L = [2,3,5,60]$.

\section{Conclusion}\label{sec:conc}

In this paper, we have considered an extension to Prolog with  
mutually exclusive  modules. This extension allows modules of 
the form  $D_0 \adc  D_1$  where $D_0, D_1$ are modules.
These modules are 
 particularly useful for structuring the program space.

  We are investigating the connection between \hweb\ and Japaridze's
computability logic \cite{Jap03,Jap08}.

\bibliographystyle{plain}


\end{document}